\documentclass[12pt]{iopart}

\usepackage{iopams}
\usepackage{graphicx}
\usepackage{bm}

\newcommand{\mean}[1]{\langle{#1}\rangle}
\newcommand{\pro}[2]{\langle{#1}|{#2}\rangle}
\newcommand{\bra}[1]{\langle{#1}|}
\newcommand{\ket}[1]{|{#1}\rangle}

\begin{document}


\title[]
{Optimal control for perfect state transfer in linear quantum memory}

\author{Hideaki Nakao and Naoki Yamamoto}

\address{
Department of Applied Physics and Physico-Informatics, 
Keio University, Hiyoshi 3-14-1, Kohoku, Yokohama 223-8522, Japan}

\ead{yamamoto@appi.keio.ac.jp}


\begin{abstract}
A quantum memory is a system that enables transfer, storage, and retrieval of 
optical quantum states by ON/OFF switching of the control signal in each 
stages of the memory. 
In particular, it is known that, for perfect transfer of a single-photon state, 
appropriate shaping of the input pulse is required. 
However, in general, such a desirable pulse shape has a complicated form, 
which would be hard to generate in practice. 
In this paper, for a wide class of linear quantum memory systems, we develop 
a method that reduces the complexity of the input pulse shape of a single-photon 
while maintaining the perfect state transfer. 
The key idea is twofold; (i) the control signal is allowed to vary continuously in 
time to introduce an additional degree of freedom, and then (ii) an optimal control 
problem is formulated to design a simple-formed input pulse and the 
corresponding control signal. 
Numerical simulations are conducted for $\Lambda$-type atomic media and 
a networked atomic ensembles, to show the effectiveness of the proposed 
method. 
\end{abstract}




\section{Introduction}

Photons, due to its nature of low interaction with the environment, is the most 
popular candidate of an information carrier for secure quantum communication. 
The technology of light transmission has been well developed, which raises its 
popularity. 
However, because a photon is not well suited for local manipulation, a photonic 
state must be converted to a stationary state of a solid system for storage 
and information processing \cite{DiVincenzo}. 
In fact, various methods for such a state conversion have been studied 
\cite{Phillips, Julsgaard, Chang, Bao}. 
In particular, the electromagnetic induced transparency (EIT) effect can be 
employed to switch the couplings between the metastable states of atoms and 
an optical field to preserve the state once the photon has been absorbed into 
the atoms \cite{Fleischhauer}. 
We refer to Refs. \cite{Lvovsky,Bussieres} for reviewing the recent progress of 
optical quantum memory.

\begin{figure}
\centering
\includegraphics[scale=0.55]{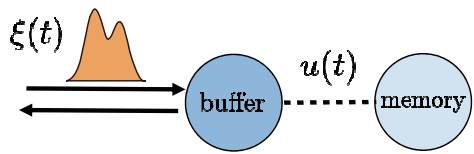}
\caption{
Configuration of a general optical quantum memory. 
The pulse function $\xi(t)$ of a single-photon field and the control signal $u(t)$ 
are properly designed so that the single photon is effectively transferred to the 
memory subsystem.}
\label{fig:QuantumMemory}
\end{figure}

Now let us consider the general configuration of a quantum memory system. 
As illustrated in Fig.~\ref{fig:QuantumMemory}, it consists of two subsystems: 
the buffer subsystem and the memory subsystem. 
The buffer subsystem is coupled to both the external field and the memory 
subsystem, while the memory subsystem is only coupled to the buffer 
subsystem. 
The interaction between the two subsystems can be controlled by an ON/OFF 
switching signal $u(t)$ so that the quantum state is preserved inside 
the memory subsystem once all the state is fed into it. 
Given a physical system having the above configuration, then the problem is 
how to design an efficient or even optimal transfer, storage, and retrieval 
processes. 
For instance, for an atomic quantum memory incorporating the EIT effect, 
we find some methods that compute a time-varying control signal $u(t)$ 
achieving optimal state transfer, based on the information gained from 
the leaking output 
\cite{Gorshkov2007,Novikova,Gorshkov,Phillips 2008,Novikova 2008}. 
Also, for a wide class of open quantum systems, i.e., {\it passive linear 
quantum systems} such as a large atomic ensemble and an opto-mechanical 
crystal array \cite{Chang,Gorshkov,Gough 2008,Guta 2016}, the general 
procedure for achieving a {\it perfect} state transfer was developed \cite{NYJ}; 
in particular, it was found that the pulse shape of an input photon, $\xi(t)$, 
must be the generalized {\it rising-exponential} function to accomplish the 
perfect state transfer if the ON signal takes a constant value. 
However, it is often the case that a generalized rising-exponential function 
has a complicated form, and experimental implementation of such a complex 
pulse shaping of a single photon field is a challenging task. 
In fact, even in the case of generating a simple rising exponential such as 
$\xi(t)=e^{t/2}$, we need to elaborate a sophisticated photon generator based 
on a cold atomic media \cite{Du 2012,Gulati PRL,Gulati PRA,Lvovsky 2015} 
or an asymmetric optical parametric oscillator \cite{Ogawa}. 
Extension of these schemes to the general multi-mode setup must be more 
involved and challenging.

The contribution of this paper is to solve the above mentioned issue; 
that is, for a general passive linear quantum system having the configuration 
shown in Fig.~\ref{fig:QuantumMemory}, we develop an optimal control 
theory for computing a low-complexity pulse function $\xi(t)$ of an input 
single-photon that can be perfectly absorbed into the memory subsystem, 
by employing a time-varying control signal $u(t)$. 
This idea is based on the fact that, in contrast to the difficulty for shaping 
a complex pulse function of a single-photon field in an experiment, there are 
many setups where even a complicated and fast control signal can be easily 
implemented using a laser pulse shaping; 
this fact has been demonstrated in the field of quantum (open-loop) optimal 
control \cite{Werschnik,Cong}. 
That is, our method can be used to make the perfect state transfer protocol 
more feasible, by converting the complexity of $\xi(t)$ (hard to implement) to 
that of $u(t)$ (easy to implement compared to $\xi(t)$). 
Note that a similar approach, but for a concrete example, was developed in 
\cite{Gorshkov2007,Novikova,Gorshkov,Phillips 2008,Novikova 2008}; 
a remarkable difference of these works to our approach is that in their setup 
the input pulse function $\xi(t)$ is {\it given} and then the optimal control 
{\it maximizing} the storage efficiency is computed; 
note thus that, in this setup, the state transfer is in general not perfect. 
Meanwhile, in our case, both the optimal pulse function and the optimal 
control signal can be found out of many combinations of those, in such 
a way that the perfect state transfer is guaranteed.

This paper is organized as follows. 
Section \ref{sec:Preliminaries} presents the general model analyzed throughout 
the paper and introduces the dynamical equations for the single-photon pulse 
function. 
In Section \ref{sec:Model}, we derive the general condition for the pulse function 
and the control signal to achieve the perfect state transfer. 
Section \ref{sec:OptimalControl} begins with the formulation of optimal control 
problem and then presents two examples, $\Lambda$-type atomic media 
and a networked atomic ensembles system. 
A concrete algorithm for solving the optimal control problem is given in 
Appendix~B.

\textbf{Notations:} 
For a matrix $A=(a_{ij})$, the symbols $A^\dagger$, $A^\top$, and 
$A^\sharp$ represent its Hermitian conjugate, transpose, and complex 
conjugation in elements of $A$, respectively; i.e., 
$A^\dagger=(a_{ji}^*)$, $A^\top=(a_{ji})$, and $A^\sharp=(a^*_{ij})$. 
For a matrix of operators we use the same notation, in which case $a_{ij}^*$ 
denotes the adjoint to $a_{ij}$. 
For a time-dependent variable $x(t)$, we denote $\dot{x}(t)=dx(t)/dt$.


\section{Preliminaries}
\label{sec:Preliminaries}


\subsection{Passive linear quantum systems}

In this work, we follow the same model presented in \cite{NYJ}, 
the general {\it passive linear quantum system} 
\cite{Chang,Gorshkov,Gough 2008,Guta 2016}. 
This is an open quantum system composed of $n$ harmonic oscillators 
represented by a vector of annihilation operators $\bm{a}=[a_1, \ldots, a_n]^T$, 
satisfying the commutation relation $a_i a_j^*-a_j^* a_i=\delta_{ij}$. 
The system interacts with a single external optical field, whose annihilation process 
is denoted by the operator $b(t)$, satisfying $b(t)b^*(s)-b^*(s)b(t)=\delta(t-s)$. 
The system has a time-varying Hamiltonian of the form 
$H(t)=\bm{a}^{\dagger}\Omega(t) \bm{a}$, where $\Omega(t)$ is an 
$n\times n$ Hermitian matrix. 
Further, the system-field instantaneous coupling is represented by 
$H_{\rm int}(t)=i[b^*(t)Ca-a^\dagger C^\dagger b(t)]$, where $C$ is a 
$n$-dimensional complex row vector. 
Then from the input-output formalism \cite{Gardiner}, the Heisenberg 
dynamics of the system variable, 
$\bm{a}(t)=U^*(t)\bm{a} U(t)=[U^*(t) a_1 U(t), \ldots, U^*(t) a_n U(t)]^T$ 
with $U(t)$ the joint unitary evolution of the system and the field, together 
with the change of the field operator, is given by:
\begin{equation}
\label{eqn:HL}
     \dot{\bm{a}}(t)=A(t)\bm{a}(t)-C^{\dagger}b(t),~~~
     \tilde{b}(t)=C\bm{a}(t)+b(t),
\end{equation}
where $A(t)=-i\Omega(t)-C^{\dagger}C/2$ and $\tilde{b}(t)$ is the annihilation 
process operator of the output field. 
Note that the vector $C$ can be time-varying, but in this paper it is assumed to 
be time-invariant.


\subsection{Single photon field}

The annihilation operator for the continuous-mode single-photon state is 
defined by 
\begin{eqnarray}
	B(\xi)=\int_{-\infty}^{\infty}\xi^*(t)b(t)dt,
\end{eqnarray}
where $\xi(t)$ is a $\mathbb{C}$-valued pulse shape function satisfying 
the normalization condition $\int_{-\infty}^{\infty}|\xi(t)|^2dt=1$ (thus 
$|\xi(t)|^2$ have the meaning of a probability distribution). 
A single photon field state is defined as 
\begin{equation}
\label{single photon state}
	|1_{\xi}\rangle=B^*(\xi)|0\rangle 
	    = \int_{-\infty}^{\infty}\xi(t)b^*(t)dt|0\rangle,
\end{equation}
where $\ket{0}$ denotes the field vacuum state. 
As expected, the operator $b(t)$ annihilates the photon; 
$b(t)\ket{1_\xi}=\xi(t)\ket{0}$. 
Note that the method proposed in this paper also functions for the case 
when the input is given by a pulsed coherent field state. 
However, for a coherent field state, even a complex wave packet can be 
effectively generated by an electro-optic modulator, and thus our method 
may not need to be applied in this case.


\subsection{Dynamics of the pulse shape and the statistics}

In this paper, the input field state for the system \eref{eqn:HL} is given by 
the single photon field state \eref{single photon state}. 
The system's initial state, at time $t=t_0$, is assumed to be the ground state 
$|\bm{0}\rangle=|0\rangle^{\otimes n}$. 
Thus, the initial state of the whole system is 
$\ket{\bm{0},1_\xi}=\ket{\bm{0}}\otimes\ket{1_\xi}$. 
Then, due to the passivity property, the system's output state on the field 
$\tilde{b}$ is also a single photon state; 
let $\tilde{\xi}(t)$ be the pulse function of this output single photon state. 
In this setup, the input pulse shape $\xi(t)$ and the output pulse shape 
$\tilde{\xi}(t)$ are connected by the following classical differential 
equation having the same form as that of Eq.~\eref{eqn:HL} \cite{NYJ,Hush,Zhang}: 
\begin{equation}
\label{eqn:IS}
      \dot{\bm{\eta}}(t)=A(t)\bm{\eta}(t)-C^{\dagger}\xi(t),~~~
      \tilde{\xi}(t)=C\bm{\eta}(t)+\xi(t).
\end{equation}
The derivation of this equation is given in Appendix~A; 
in particular, the definition of the $\mathbb{C}^n$-valued vector $\bm{\eta}(t)$ 
is given in Eq.~\eref{def of eta}. 
At time $t=t_1$ when the perfect state transfer is achieved, $\bm{\eta}(t_1)$ 
coincides with the superposition coefficients of the system state (see Appendix~A); 
this will be used as a terminal condition of the optimization problem formulated 
in Section~4. 
Note that, if the input is a coherent field state with amplitude $\xi(t)$, we have 
the same input-output relationship as above, in which case $\bm{\eta}(t)$ stands 
for the vector of means of the system coherent state.

Next let us define the correlation matrix as 
$\langle N\rangle=(\langle\bm{0},1_{\xi}|a^*_ia_j|\bm{0},1_{\xi}\rangle)$, 
whose diagonal terms represent the mean photon number in each mode. 
The dynamics of $\langle N\rangle$ is given by
\begin{equation}
\label{eqn:stateq}
     \frac{d}{dt}\langle N\rangle
         =A^{\sharp}\langle N\rangle + \langle N\rangle A^T
            -\xi^*(t)C^T \bm{\eta}^T
                -\xi(t) \bm{\eta}^{\sharp}C^{\sharp},
\end{equation}
where $\bm{\eta}(t)$ obeys Eq.~\eref{eqn:IS}.


\section{Condition for perfect state transfer}
\label{sec:Model}


\subsection{Linear quantum memory system}

\begin{figure}
\centering
\includegraphics[scale=0.5]{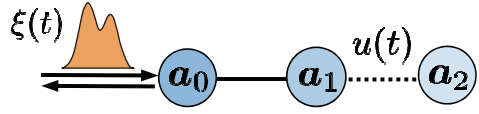}
\caption{
\label{fig:toy}
The relation between the three groups of modes; 
$(a_0, \bm{a}_1)$ denotes the buffer subsystem and $\bm{a}_2$ does the 
memory subsystem. 
The arrow represents the input and the output fields. 
The solid line represents the regular interaction, and the dotted line 
represents the tunable interaction 
with strength $u(t)$.}
\end{figure}

Throughout this paper, we consider a particular class of passive linear 
quantum systems which has the ability for transfer, storage, and retrieval of 
a quantum state. 
We let the system to have three components 
$\bm{a}=[a_0, \bm{a}_1^T, \bm{a}_2^T]^T$, where $a_0$ is a single 
annihilation operator, and $\bm{a}_1$ and $\bm{a}_2$ are vectors of 
$n_1$ and $n_2$ annihilation operators, respectively. 
Then the Hamiltonian matrix $\Omega(t)$ is given by $\Omega(t)=F+Gu(t)$, 
where 
\begin{eqnarray}
     F=\left[ \begin{array}{ccc}
             F_{00} & F_{01} & \bm{0}^T\\
             F_{01}^{\dagger} & F_{11} & O\\
             \bm{0} & O & O
          \end{array}\right],~~
      G=\left[\begin{array}{ccc}
              0 & \bm{0}^T & \bm{0}^T\\
              \bm{0} & G_{11} & G_{12}\\
              \bm{0} & G_{12}^{\dagger} & G_{22}
           \end{array}\right],
\nonumber
\end{eqnarray}
are time-invariant matrices and $u(t)\in\mathbb{R}$ is a control signal which 
modulates the Hamiltonian. 
The size of the matrices $F_{ij}$ and $G_{ij}$ is $n_i\times n_j$, where 
we have defined $n_0=1$. 
We also define $C=[ c, \bm{0}^T, \bm{0}^T]$ with $c\in{\mathbb C}$. 
Hence, Eq. \eref{eqn:HL} is now given by 
\begin{eqnarray}
\hspace{-2cm}
         \frac{d}{dt}
	\left[\begin{array}{c}
		a_0\\
		\bm{a}_1\\
		\bm{a}_2
	\end{array}\right]
       &=\left[\begin{array}{ccc}
                  -\frac{1}{2}|c|^2-iF_{00} & -iF_{01} & \bm{0}^T\\
                  -iF_{01}^{\dagger} & -iF_{11}-iG_{11}u(t) & -iG_{12}u(t)\\
                  \bm{0} & -iG_{12}^{\dagger}u(t) & -iG_{22}u(t)
            \end{array}\right]
            \left[\begin{array}{c}
                  a_0\\
                  \bm{a}_1\\
                  \bm{a}_2
	\end{array}\right]
        -\left[\begin{array}{c}
                  c^*\\
                  \bm{0}\\
                  \bm{0}
         \end{array}\right]b,
\nonumber \\ & \hspace{-0.2cm}
          \tilde{b}=ca_0+b.
\label{eqn:arg}
\end{eqnarray}
It is immediate to see that, during $u(t)=0$, the second mode $\bm{a}_2$ 
entirely decouples from the input field and the other system modes. 
Therefore, $\bm{a}_2$ functions as a memory subsystem, because once the 
photon is transferred into the state of $\bm{a}_2$, that state can be preserved 
for a long time by tuning $u(t)=0$. 
The other modes $a_0$ and $\bm{a}_1$ are the buffer subsystems, which 
mediate the state transfer from the optical field to the memory subsystem. 
The relation of the three groups of modes is illustrated in Fig.~\ref{fig:toy}. 
Finally note that the input and output pulse functions $\xi(t)$ and 
$\tilde{\xi}(t)$ are connected by Eq.~\eref{eqn:IS}, which is now 
\begin{eqnarray*}
\hspace{-2cm}
         \frac{d}{dt}
         \left[\begin{array}{c}
                 \eta_0\\
                 \bm{\eta}_1\\
                 \bm{\eta}_2
         \end{array}\right]
       &=\left[\begin{array}{ccc}
                  -\frac{1}{2}|c|^2-iF_{00} & -iF_{01} & \bm{0}^T\\
                  -iF_{01}^{\dagger} & -iF_{11}-iG_{11}u(t) & -iG_{12}u(t)\\
                  \bm{0} & -iG_{12}^{\dagger}u(t) & -iG_{22}u(t)
            \end{array}\right]
            \left[\begin{array}{c}
                  \eta_0\\
                  \bm{\eta}_1\\
                  \bm{\eta}_2
           \end{array}\right]
        -\left[\begin{array}{c}
                  c^*\\
                  \bm{0}\\
                  \bm{0}
         \end{array}\right]\xi,
\nonumber \\ & \hspace{-0.2cm}
          \tilde{\xi}=c\eta_0+\xi.
\end{eqnarray*}

\subsection{Zero-dynamics principle for perfect state transfer}
\label{sec:PT3}

For the perfect transfer of a photonic state, the output pulse function must 
always be zero during the state transfer process; 
this is the idea of zero-dynamics principle for perfect state transfer \cite{NYJ}. 
Thus, from Eq.~\eref{eqn:IS}, the condition we desire is 
$\tilde{\xi}(t)=c\eta_0(t)+\xi(t)=0$. 
The time derivative of the output pulse is then calculated as 
\[
\hspace{-1cm}
     \dot{\tilde{\xi}}(t) = c\,\dot{\eta}_0(t)+\dot{\xi}(t)
     = c\left(-\frac{1}{2}|c|^2-iF_{00}\right)\eta_0(t) - iF_{01}c\,\bm{\eta}_1(t)
            -c^*c\,\xi(t) + \dot{\xi}(t),
\]
which is also zero since the output must be constantly zero. 
Substituting the zero-output condition $\eta_0(t)=-\xi(t)/c$ for the above 
equation, we have
\begin{equation}
\label{eqn:xieq}
\hspace{-1.2cm}
	\frac{d}{dt}\left[
	\begin{array}{c}
		\xi(t)\\
		\bm{\eta}_1(t)\\
		\bm{\eta}_2(t)
	\end{array}
	\right]=\left[
	\begin{array}{ccc}
		\frac{1}{2}|c|^2-iF_{00} & iF_{01}c & \bm{0}^T\\
		iF_{01}^{\dagger}/c & -iF_{11}-iG_{11}u(t) & -iG_{12}u(t)\\
		\bm{0} & -iG_{12}^{\dagger}u(t) & -iG_{22}u(t)
	\end{array}
	\right]\left[
	\begin{array}{c}
		\xi(t)\\
		\bm{\eta}_1(t)\\
		\bm{\eta}_2(t)
	\end{array}
	\right].
\end{equation}
This equation can be simply represented as
\begin{equation}
\label{eqn:xieq 2}
	\dot{\bm{x}}(t)=\left(A_0+A_1u(t)\right)\bm{x}(t),
\end{equation}
where $\bm{x}=[\xi, \bm{\eta}^T_1, \bm{\eta}^T_2]^T$ and $(A_0, A_1)$ 
are fixed matrices. 
Let us next derive the boundary condition from the system's final state when 
the state transfer is accomplished at time $t_1$. 
First, since $\bm{\eta}(t_1)$ coincides with the terminal superposition coefficients 
of $\bm{a}(t_1)$, we have $\eta_0(t_1)=0$ and $\bm{\eta}_1(t_1)=\bm{0}$ 
for the perfect state transfer into the memory subsystem. 
Also from $\xi(t)=-c\eta_0(t)$, the input pulse must satisfy $\xi(t_1)=0$. 
In total, $\bm{x}(t_1)=[0, \bm{0}^T, \bm{\eta}^T_2(t_1)]^T$. 
The dynamics \eref{eqn:xieq} or \eref{eqn:xieq 2} with this boundary condition 
is called the zero-dynamics, and it can be uniquely solved backwards in time 
starting from $t_1$ once $u(t)$ is specified. 
In other words, this dynamics is the condition imposed on $u(t)$ and $\xi(t)$ 
for achieving the perfect state transfer.


\subsection{Constant control and rising exponential function}

Let us consider the special case when the control signal $u(t)$ is constant. 
In this case, letting $\tilde{\xi}(t)=0$ in Eq.~\eref{eqn:IS} readily yields 
\begin{eqnarray}
\label{eqn:generalexp}
      \xi(t)=-\bm{\eta}^T(t_1) e^{A^{\sharp}(t_1-t)}C^T\Theta(t_1-t),
\end{eqnarray} 
where $\Theta(t)$ is a Heaviside step function taking $1$ for $t\geq 0$ 
and $0$ for $t<0$. 
Note of course that Eq.~\eref{eqn:generalexp} is the solution of 
Eq.~\eref{eqn:xieq} when $u(t)$ is constant.

To see explicitly a feature of the function \eref{eqn:generalexp}, 
let us consider a single-mode system specified by $\Omega=0$ and 
$C=\sqrt{\gamma}$; 
the dynamical equation of the system annihilation operator $a(t)$ is given by 
\[
     \dot{a}(t)=-\frac{\gamma}{2}a(t)-\sqrt{\gamma}b(t),~~~
     \tilde{b}(t)=\sqrt{\gamma}a(t)+b(t). 
\]
Typically, this system can be implemented as an optical cavity coupled to an 
external field, where $\gamma$ is proportional to the transmissivity of the 
coupling mirror. 
In this case, Eq.~\eref{eqn:generalexp} is given by 
$\xi(t)=-\eta(t_1) \, \sqrt{\gamma}\, e^{\gamma t/2}\Theta(-t)$, 
where $t_1=0$ is assumed for simplicity. 
That is, the pulse function is of the rising exponential form
\footnote{
A single-photon field state with this pulse function can be perfectly absorbed 
into the cavity. 
Note however that, if we want to use this system as a memory device, the 
coupling strength $\gamma$ has to be a controllable time-varying parameter; 
that is, $\gamma$ must be changed to zero immediately after the state 
transfer is finished. 
}.
Based on this observation, we call Eq.~\eref{eqn:generalexp} the generalized 
rising exponential function, for systems with the number of modes more than 
or equal to two. 
In fact, in terms of the eigenvalues of $A$, $\{\lambda_i\}$, 
Eq.~\eref{eqn:generalexp} can be represented as 
$\xi(t)=\sum_{i=1}^n \xi_i e^{-\lambda^*_i t}$ with $\xi_i$ the constant 
coefficients; 
hence, under the assumption that ${\rm Re}(\lambda_i)<0$, which is required 
to achieve the perfect state transfer \cite{NYJ}, Eq.~\eref{eqn:generalexp} 
is indeed the sum of rising exponential functions. 
Note again that a single-photon field state over the pulse function 
\eref{eqn:generalexp} is perfectly absorbed into the corresponding 
multi-mode passive linear system \eref{eqn:HL}. 
Also we remark that Eq.~\eref{eqn:generalexp} is the time-reversed version 
of the system's output field \cite{NYJ}, as in the single-mode case 
\cite{Leuchs 2000,Bader}. 
However, as we refer in the examples in the next section, a generalized 
rising exponential function has a complicated form in general; 
thus realization of a single-photon field state with such a complex pulse function 
is still a challenging task, while, as mentioned in Section~1, a simple rising 
exponential of a single-photon field state is possible to experimentally generate, 
using current technology.


\section{Optimal control for quantum memory}
\label{sec:OptimalControl}

The main purpose for considering the variation of the control signal $u(t)$ 
is to reduce the complexity of the pulse function \eref{eqn:generalexp}, 
while maintaining the perfect state transfer. 
The problem is that, as implied by Eq.~\eref{eqn:xieq}, there are infinitely 
many combinations of $\xi(t)$ and $u(t)$ that satisfy the condition for 
perfect state transfer. 
In this section, we present an optimal-control formulation for determining 
the optimal pair of low-complexity pulse function and the corresponding 
control signal, and then show two examples.


\subsection{Designing the cost function}

As mentioned above, the general rising-exponential pulse function of a 
single-photon is often of a very complicated form, which would be hard to 
generate. 
On the other hand, a single-photon field traveling with a simple rising exponential 
pulse function such as $\xi(t)=-\eta(t_1) \, \sqrt{\gamma}\, e^{\gamma t/2}\Theta(-t)$ 
shown above can be experimentally generated 
\cite{Du 2012,Gulati PRL,Gulati PRA,Lvovsky 2015,Ogawa}, although such 
a simple function cannot perfectly transfer the photon into the memory subsystem
\footnote{
From Eq.~\eref{eqn:xieq} and the boundary condition, the time derivative 
of the pulse function at time $t_1$ is $\dot{\xi}(t_1)=0$. 
Thus, the simple rising exponential function does not transfer the state into 
the memory subsystem perfectly, except the case where the system is a 
single-mode system; see the footnote in Page~6. 
}. 
Therefore, as a compromise, we consider a unimodal function; 
that is, here we formulate an optimal control problem such that the cost 
function is minimized when the pulse function takes a unimodal form. 
This policy is supported by the fact that a source system generating a single 
photon with unimodal pulse function has been experimentally implemented 
in several setups, such as cavity QED \cite{Rempe2002,Kuhn2011,Rempe2012}, 
ion trap \cite{Hayasaka2004}, circuit QED \cite{Filipp2014}, and cold atoms 
\cite{Harris2008,Riedmat2016}. 
Actually, by properly modulating the source system during the photon emission 
process, we can flexibly tailor the pulse shape so that the desired unimodal 
pulse function is realized. 
For instance, it was demonstrated in the ion-trap setup in \cite{Hayasaka2004} 
that the duration of the Gaussian-shaped pulse of the emitted single photon 
can be controlled by tuning the intensity of the driving field for the ion; 
also, the circuit-QED-based scheme developed in \cite{Filipp2014} generates 
a single photon with various $\sin^2$-shaped form, by a proper modulation 
of the driving microwave signal that effectively changes the qubit-resonator 
coupling.

\begin{figure}[tb]
\begin{center}
\includegraphics[width=0.62\textwidth]{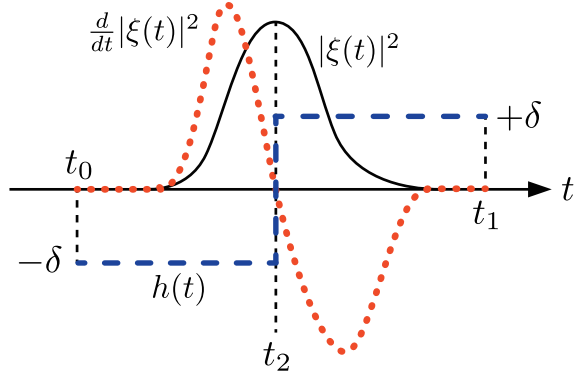}
\caption{
Relation between $|\xi(t)|^2$ (represented by the solid black line), 
$d|\xi(t)|^2/dt$ (the red dotted line), and $h(t)$ (the blue dashed line). 
}
\label{fig:Pulsecost}
\end{center}
\end{figure}

According to the above-described policy, we define a cost function so that 
$|\xi(t)|^2$ would monotonically increase until it meets its single extremum 
and decrease after the pulse reaches that extremum. 
Let $t_2$ be the time of this extremum; hence $t_0<t_2<t_1$. 
Then, the cost function should impose a penalty when the pulse function is not 
increasing until $t_2$ and not decreasing after $t_2$. 
A simple optimal control problem satisfying this policy can be defined as 
\[
     \min_{\{u(t)\}}\int_{t_0}^{t_1} h(s) \frac{d}{ds}|\xi(s)|^2ds,
\]
where 
\[
     h(t) = \left\{ \begin{array}{ll}
                  -\delta  &  (t_0 \leq t < t_2) \\
                  +\delta &  (t_2 \leq t < t_1) \\
               \end{array} \right.,
\]
for a dimensionless constant $\delta>0$. 
Actually this cost function takes a smaller value if $d|\xi(t)|^2/dt>0$ 
for $t_0\leq t\leq t_2$ and $d|\xi(t)|^2/dt<0$ for $t_2\leq t \leq t_1$; 
Fig.~\ref{fig:Pulsecost} illustrates the relation between $|\xi(t)|^2$, 
$d|\xi(t)|^2/dt$, and $h(t)$ in this desired case. 
Note that $t_2$ is a decision variable. 
Also the constraint of this optimization problem is given by Eq.~\eref{eqn:xieq 2}, 
which together with the boundary condition is 
\begin{eqnarray}
        \dot{\bm{x}}(t)&=\left(A_0+A_1u(t)\right)\bm{x}(t),
\nonumber\\
        \bm{x}(t_1)
             &=\left[\begin{array}{ccc}
                         0, & \bm{0}^T, & \bm{\eta}_2^T(t_1)
	         \end{array}\right]^T.
\label{eqn:subjectsimple}
\end{eqnarray}

The above optimal control problem can be generalized to 
\begin{eqnarray}
          \min_{\{u(t)\}}
             \int_{t_0}^{t_1}f\left(h(s)\frac{d}{ds}|\xi(s)|^2\right)ds,
\label{eqn:cost1}
\end{eqnarray}
where $f(x)$ is an increasing function introduced to enhance the penalty. 
In particular, we set $f(x)=e^x$. 
Moreover, the cost over the control sequence and the boundary condition 
should be also taken into account. 
Thus, the overall cost function is given by 
\begin{equation}
\label{eqn:finalcostfunc}
      J[u] = \int_{t_0}^{t_1}
          \left[\alpha \, {\rm exp}\left(h(s)\frac{d}{ds}|\xi(s)|^2\right) 
                     + \beta u^2(s)\right]ds +\gamma \|\bm{x}(t_0)\|^2.
\end{equation}
Here, $(\alpha,\beta,\gamma)$ are dimensionless positive constants that 
change the weight of each cost. 
Summarizing, the optimal control problem is to find $u(t)$ that minimizes the 
cost \eref{eqn:finalcostfunc} under the constraint \eref{eqn:subjectsimple}. 
Because the variable is fixed at the terminal time $t_1$, this optimization 
problem is solved backwards in time; 
the initial time $t_0$ is chosen so that the initial system state is enough close 
to the ground state, or equivalently that the initial variable $\bm{x}(t_0)$ is 
enough close to zero. 
The detailed procedure for calculating the optimal pair of $\xi(t)$ and $u(t)$ 
is given in Appendix~B.



\subsection{$\Lambda$-type atomic media}

\begin{figure}[tb]
\begin{center}
\includegraphics[scale=0.5]{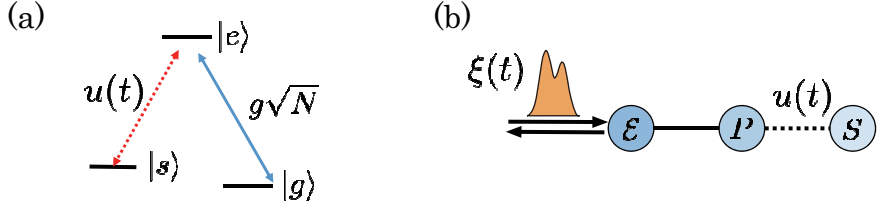}
\caption{
(a) The energy diagram of a $\Lambda$-type atom. 
(b) Relation of the modes; $(\mathcal{E}, P)$ is the buffer subsystem and 
$S$ is the memory subsystem.}
\label{fig:schematom}
\end{center}
\end{figure}

In this section, we study a $\Lambda$-type atomic ensemble trapped in an 
optical cavity, investigated in \cite{Gorshkov2007,Novikova,Gorshkov,
Phillips 2008,Novikova 2008}. 
In this model, there are $N$ atoms with three energy states as shown in 
Fig.~\ref{fig:schematom}~(a), where the $|g\rangle \leftrightarrow |e\rangle$ 
transition of frequency $\omega_{eg}$ couples to the cavity radiation mode 
with frequency $\omega_1$, and the $|s\rangle \leftrightarrow |e\rangle$ 
transition of frequency $\omega_{es}$ couples to an external control field with 
frequency $\omega_2$ having a time-varying Rabi frequency envelope 
$u(t)\in{\mathbb R}$. 
The system variables are the polarization operator $P(t)=\sigma_{ge}(t)/\sqrt{N}$, 
the spin-wave operator $S(t)=\sigma_{gs}(t)/\sqrt{N}$ ($\sigma_{\mu\nu}$ are 
collective atomic operators for $|\mu\rangle\langle\nu|$), and the cavity 
mode operator $\mathcal{E}(t)$. 
When $N\gg1$, the operators $P$ and $S$ can be approximated as 
annihilation operators, and the Heisenberg equation of $(\mathcal{E},P,S)$ 
is given by 
\[
         \frac{d}{dt}
         \left[\begin{array}{c}
              \mathcal{E}(t)\\
              P(t)\\
              S(t)
         \end{array}\right] 
        = \left[\begin{array}{ccc}
                  -\kappa & ig\sqrt{N} & 0\\
                  ig\sqrt{N} & -i\Delta & i u(t)\\
                 0 & i u(t) & 0
            \end{array}\right]
            \left[\begin{array}{c}
                 \mathcal{E}(t) \\
                 P(t) \\
                 S(t) \\
            \end{array}\right]
            + \left[\begin{array}{c}
                 \sqrt{2\kappa} \\
                 0 \\
                 0 \\
              \end{array} \right]b(t),
\]
where $g\in\mathbb{R}$ is a coupling constant between the atoms and the 
cavity field. 
Also $\Delta=\omega_{eg}-\omega_1=\omega_{es}-\omega_2$ denotes 
the detuning, which is assumed to be zero for simplicity. 
The cavity decay rate is $2\kappa$, so the relation between the input process 
operator $b(t)$ and the output correspondence $\tilde{b}(t)$ is given by 
\begin{eqnarray}
        \tilde{b}(t)=\sqrt{2\kappa}\,\mathcal{E}(t)-b(t).
\end{eqnarray}
Now defining $b'(t)=-ib(t)$ and $\tilde{b}'(t)=i\tilde{b}(t)$, the above 
equations can be represented in the form of passive linear quantum system 
\eref{eqn:HL} with system matrices 
\[
        A(t) = \left[\begin{array}{ccc}
                   -\kappa & ig\sqrt{N} & 0\\
                   ig\sqrt{N} & 0 & i u(t)\\
                   0 & i u(t) & 0
                \end{array}\right],~~~
           C = \left[\begin{array}{ccc}
                         i\sqrt{2\kappa} & 0 & 0
                     \end{array}\right],
\]
and the system variables $\bm{a}= [a_1, a_2, a_3]^T = [\mathcal{E}, P, S]^T$. 
That is, $(a_1, a_2)=(\mathcal{E}, P)$ functions as the buffer subsystem and 
$a_3=S$ does the memory subsystem, as illustrated in 
Fig.~\ref{fig:schematom}~(b); 
actually when $u(t)=0$, the spin-wave mode $S$ is decoupled from the other 
modes and the field mode $b(t)$. 
Note that the parameters defined in Eq.~\eref{eqn:arg} are $c=i\sqrt{2\kappa}$, 
$F_{00}=0$, $F_{01}=-g\sqrt{N}$, $G_{11}=0$, $G_{12}=-1$, and $G_{22}=0$.

\begin{figure}[tb]
\centering
\includegraphics[scale=0.7]{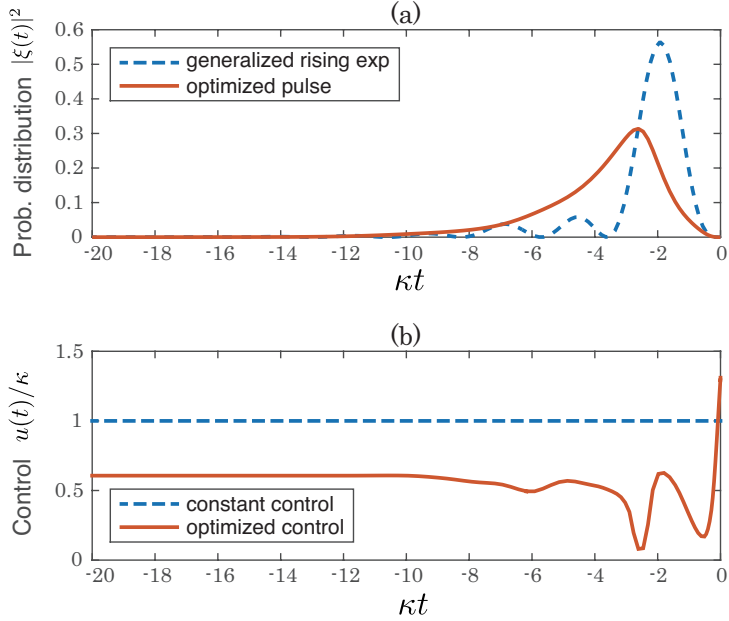}
\caption{
(a) Input pulse function $|\xi(t)|^2$; the generalized rising exponential 
corresponding to the constant control (dashed blue line) and the optimized 
unimodal function corresponding to the time-varying control (solid red line). 
(b) The constant control signal $u(t)=\kappa \,\Theta(-t)$ (dashed blue line), 
and the optimized time-varying control signal (solid red line).}
\label{fig:pulseshape}
\end{figure}

Now the objective is to transfer a single-photon to the spin-wave mode $S$, 
implying that the final state is set to $\bm{\eta}(t_1)=[0, 0, 1]^T$. 
The pair of optimal low-complexity input pulse function $\xi(t)$ and the 
corresponding control signal $u(t)$, which achieves the perfect state transfer, 
can be computed by solving the optimal control problem developed in the 
previous subsection; 
more precisely, as described in Appendix~B, the gradient method 
\cite{Bryson Ho,Nocedal} combined with the Euler-Lagrange equations 
\eref{eqn:fx}, \eref{eqn:adjfx}, and \eref{eqn:dHdu} subject to the 
dynamics \eref{eqn:subjectto} and the cost function \eref{eqn:finalcostfunc} 
yields the optimal solution. 
The system parameters are chosen so that $\kappa=g\sqrt{N}$. 
Also the dimensionless weighting parameters are set to 
$(\alpha, \beta, \gamma, \delta)=(10,1,10^4,20)$; 
the reason of choosing a large value of $\gamma$ is to strongly impose the 
initial variable $\bm{x}(t_0)$ to be close to zero.

Figure~\ref{fig:pulseshape}~(a) shows the absolute value of pulse function 
(probability density), i.e., $|\xi(t)|^2$ with $t_0=-20/\kappa$ and $t_1=0$. 
The dashed blue line represents the generalized rising exponential function 
\eref{eqn:generalexp} corresponding to the constant control 
$u(t)=\kappa \,\Theta(-t)$, 
and the solid red line shows the optimized pulse function corresponding to the 
optimal time-varying control signal $u(t)$; 
these control signals are illustrated in Fig.~\ref{fig:pulseshape}~(b). 
A remarkable fact is that, in contrast to the complicated pulse shape realized 
with the constant control, the optimized pulse function has a simple unimodal 
shape with extremum point $t_2\approx -2.6/\kappa$, thanks to the aid of 
optimal time-varying control signal. 
As discussed in Section~4.1, implementing this unimodal pulse of a single 
photon may be feasible with current technology. 
Also, the phase of the single-photon pulse is fixed to ${\rm arg}\,\xi(t)=\pi/2$ 
for all $t\leq 0$, implying that the phase modulation of the single-photon field 
is unnecessary.

Note that this time-varying signal $u(t)$ looks complicated as well, but 
the external optical field with this level of frequency envelope is feasible 
to implement, as shown in the experiment \cite{Phillips 2008,Novikova 2008}; 
recall that, on the other hand, implementing the complex pulse shape $\xi(t)$ 
corresponding to the constant control $u(t)=\kappa \,\Theta(-t)$ may be a 
challenging task.

\begin{figure}[tb]
\centering
\includegraphics[scale=0.7]{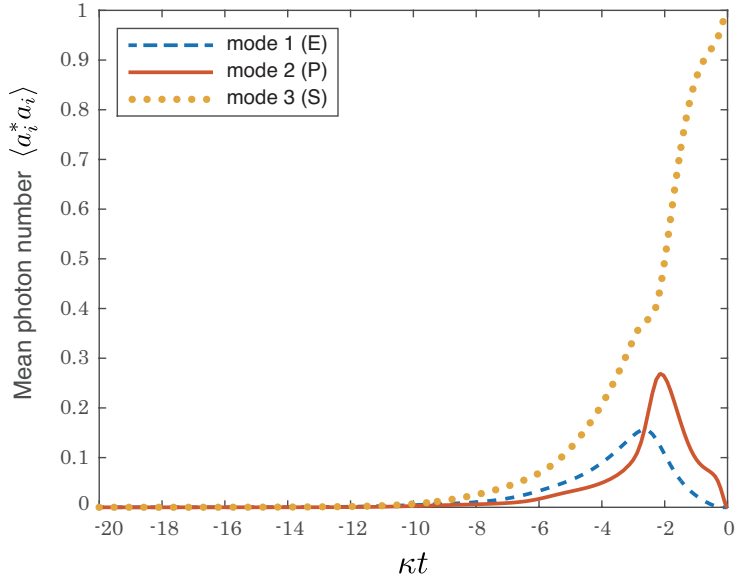}
\caption{The mean photon number in each mode. 
At time $t_1=0$, the single photon is completely transferred into the 
memory subsystem mode $a_3=S$ 
(yellow dotted line).}
\label{fig:statfigure}
\end{figure}

Finally let us confirm that the perfect state transfer is achieved. 
Figure~\ref{fig:statfigure} shows the time evolution of the mean 
photon numbers inside the atomic medium, i.e., 
$\mean{a_i^*a_i}=\bra{\bm{0},1_{\xi}}a^*_ia_i\ket{\bm{0},1_{\xi}}$, 
which can be calculated using Eq.~\eref{eqn:stateq}. 
We then actually observe that the photon has been transferred into the 
memory subsystem mode $a_3=S$ at the final time $t_1=0$, while the 
buffer subsystem $(a_1, a_2)=(\mathcal{E}, P)$ is excited during the 
time-evolution but finally decays to the ground states.

\subsection{Network of atomic ensembles}
\label{subsec:Atomic ensembles}

First of all, let us remember the fact that the optimization problem formulated 
in Section~4.1 does not necessarily yields a unimodal pulse function $\xi(t)$ 
as the optimal solution. 
Hence, it should be worth investigating to see if a unimodal input pulse 
function could be obtained for a more complicated system than the previous 
$\Lambda$-type atomic media where actually the optimal solution is given 
by a unimodal function. 
In this subsection, we study a networked system composed of large atomic 
ensembles and an optical cavity field, provided in \cite{NYJ,Duan,Parkins,Ficek}; 
in fact, as will be shown later, this system forms a target memory subsystem 
in a nontrivial way, unlike the previous example.

\begin{figure}[tb]
\begin{center}
\includegraphics[scale=0.45]{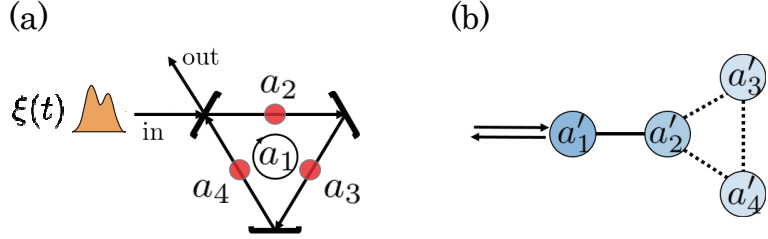}
\caption{
(a) Configuration of the system, composed of atomic ensembles and an optical 
ring cavity. 
(b) Relation of the modes; $(a_1', a_2')$ is the buffer subsystem and 
$(a_3', a_4')$ is the memory subsystem.}
\label{fig:schem}
\end{center}
\end{figure}

The configuration of the system is shown in Fig.~\ref{fig:schem}~(a). 
The network contains three atomic ensembles; 
the $k$th ensemble couples to the common single cavity mode $a_1$ through 
external pulse lasers with Rabi frequencies $\omega_k$ and $\omega'_k$. 
The coupling Hamiltonian is given by 
\begin{eqnarray} 
    H_{\rm ac}
       =\frac{\sqrt{N}\mu}{2\delta}
           \sum_{k=2}^4 \Big[ a_1^* (\omega_k e^{i\phi_k} a_k 
                                                      + \omega'_k e^{i\phi'_k}a_k^*) + \mathrm{H.c.} 
           \Big], 
\end{eqnarray}
where $a_k~(k=2,3,4)$ is the bosonic annihilation operator approximating the 
atomic collective lowering operators in the large ensemble limit. 
Also, $\phi_k \in [0,2\pi)$ is the laser phase, $N$ is the number of atoms in 
each ensemble, $\mu$ is the coupling strength, and $\delta$ is the detuning. 
The spontaneous emission of each atom is negligible for typical atoms 
such as $^{87}$Rb. 
Moreover, for the second and third ensembles, the ground state is coupled to 
the excited state, through another controllable laser field with Rabi frequency 
$u(t)$. 
This effect can be represented by the Hamiltonian 
$H_{\rm a}(t) = u(t) a_2^*a_2 - u(t) a_3^*a_3$; 
note that this is the sum of local Hamiltonians acting on each atomic ensemble. 
We set the parameters as $\omega_k=\omega>0,~\omega'_k=0$ and 
$\phi_k=\pi/2$ for $k=2,3,4$, and define $g=\sqrt{N}\mu \omega/2\delta$. 
As a result, the whole system Hamiltonian is given by 
\begin{eqnarray}
& & \hspace*{-5em}
    H(t) = H_{\rm a}(t) + H_{\rm ac} 
        = u(t) a_2^*a_2 - u(t) a_3^*a_3 
         + ig a_1^*(a_2+a_3+a_4) - ig (a_2^*+a_3^*+a_4^*)a_1
\nonumber \\ & & \hspace*{-3em}
       = [a_1^*, a_2^*, a_3^*, a_4^*]
           \left[ \begin{array}{cccc}
               0 & ig & ig & ig  \\
               -ig & u(t) & 0 & 0  \\
               -ig & 0 & -u(t) & 0  \\
               -ig & 0 & 0 & 0  \\
           \end{array} \right]
           \left[ \begin{array}{c}
               a_1 \\
               a_2 \\
               a_3 \\
               a_4 \\
           \end{array} \right]
        = a^\dagger \Omega(t) a.
\nonumber
\end{eqnarray}
The cavity mode interacts with an external light field $b(t)$ at the partially 
reflective mirror, via the Hamiltonian 
$H_{\rm int}(t)=i\sqrt{\kappa}[b^*(t)a-a^* b(t)]$. 
Summarizing, the system is a passive linear quantum system with the 
following system matrices:
\[
           A(t)=\left[\begin{array}{cccc}
                    -\kappa/2 & g & g & g \\
                    -g & -iu(t) & 0 & 0\\
                    -g & 0 & iu(t) & 0\\
                    -g & 0 & 0 & 0
                \end{array}\right],~~~
            C=\left[\begin{array}{cccc}
                 -\sqrt{\kappa} & 0 & 0 & 0
                 \end{array}\right].
\]
Now let us define the unitary matrix 
\[
          U=\left[\begin{array}{cccc}
                    1 & 0 & 0 & 0\\
                    0 & 1/\sqrt{3} & 2/\sqrt{6} & 0\\
                    0 & 1/\sqrt{3} & -1/\sqrt{6} & 1/\sqrt{2}\\
                    0 & 1/\sqrt{3} & -1/\sqrt{6} & -1/\sqrt{2}\\
                \end{array}\right],
\]
which transforms the system equations to
\[
         \dot{\bm{a}}'(t)=A'(t)\bm{a}'(t)-C'^{\dagger}b(t),~~~
         \tilde{b}(t)=C'\bm{a}'(t)+b(t),
\]
where
\begin{eqnarray}
       & \bm{a}'
         =U^{\dagger}\bm{a}
         =\left[\begin{array}{c}
                a_1\\
                (a_2+a_3+a_4)/\sqrt{3}\\
                (2a_1-a_3-a_4)/\sqrt{6}\\
                (a_3-a_4)/\sqrt{2}
          \end{array}\right],
\nonumber\\
         & A'(t)=U^{\dagger}A(t)U
               =\left[\begin{array}{cccc}
                    -\kappa/2 & \sqrt{3}g & 0 & 0\\
                    -\sqrt{3}g & 0 & -iu(t)/\sqrt{2} & iu(t)/\sqrt{6} \\
                    0 & -iu(t)/\sqrt{2} & -iu(t)/2 & -iu(t)/2\sqrt{3}\\
                    0 & iu(t)/\sqrt{6}& -iu(t)/2\sqrt{3} & iu(t)/2
                 \end{array}\right],
\nonumber\\
        & C'=CU=\left[\begin{array}{cccc}
              \sqrt{\kappa} & 0 & 0 & 0
           \end{array}\right].
\nonumber
\end{eqnarray}
This equator clearly shows that $(a'_1, a'_2)$ corresponds to the buffer 
subsystems and $(a'_3, a'_4)$ does the memory subsystems; 
once the photon is transferred to the memory subsystem, then by setting the 
control signal to $u(t)=0$, the state can be preserved. 
A striking feature of this system is that the buffer-memory interaction 
can be regulated by the local control on the second and third ensembles.

\begin{figure}[tb]
\centering
\includegraphics[scale=0.7]{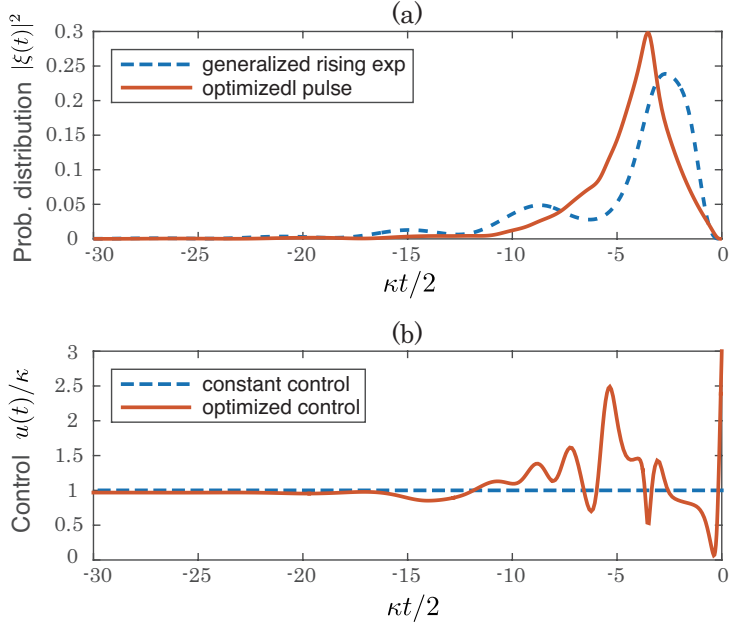}
\caption{
(a) Input pulse function $|\xi(t)|^2$; the generalized rising exponential 
corresponding to the constant control (dashed blue line) and the optimized 
unimodal function corresponding to the time-varying control (solid red line). 
(b) The constant control signal $u(t)=\kappa \,\Theta(-t)$ (dashed blue line), 
and the optimized time-varying control signal (solid red line). 
}
\label{fig:pulsecontrol}
\end{figure}

Here we set the goal to transfer a single-photon into the mode $a'_4$, implying 
that the terminal condition is given by $\bm{\eta}'(t_1)=[0, 0,0, 1]^T$. 
The parameters are chosen so that $g=\kappa/2$. 
Also the dimensionless weighting parameters are chosen as 
$(\alpha, \beta, \gamma, \delta)=(100,0.1,10^4,20)$; 
thus, the penalty on the control is taken to be smaller than the case of previous 
example. 
The initial and final time are taken to be $t_0=-60/\kappa$ and $t_1=0$. 
Figure~\ref{fig:pulsecontrol}~(a) shows the comparison of the pulse function 
$|\xi(t)|^2$ when the control signal is constant $u(t)=\kappa \,\Theta(-t)$ 
and when $u(t)$ is optimized. 
Figure~\ref{fig:pulsecontrol}~(b) illustrates those control signals. 
Clearly, the optimized pulse shape realized with the aid of time-varying control 
signal is simpler and thus more feasible, than the generalized rising exponential 
function. 
A notable fact is that the optimized pulse has a unimodal form, despite that 
the system is more involved than the previous example; 
hence we obtain a positive answer to the question we had in the beginning 
of this subsection. 
On the other hand, the control signal $u(t)$ has to vary in a complicated form 
in time, as shown in the solid red line in Fig.~\ref{fig:pulsecontrol}~(b). 
However, recall now that $u(t)$ is the time-varying coefficient of the {\it local} 
Hamiltonian $H_{\rm a}(t)$; 
hence even such a complex modulation of $u(t)$ is more feasible than the case of 
generating the generalized rising exponential $\xi(t)$ shown in the 
dashed-blue line in Fig.~\ref{fig:pulsecontrol}~(a).

\begin{figure}[tb]
\centering
\includegraphics[scale=0.7]{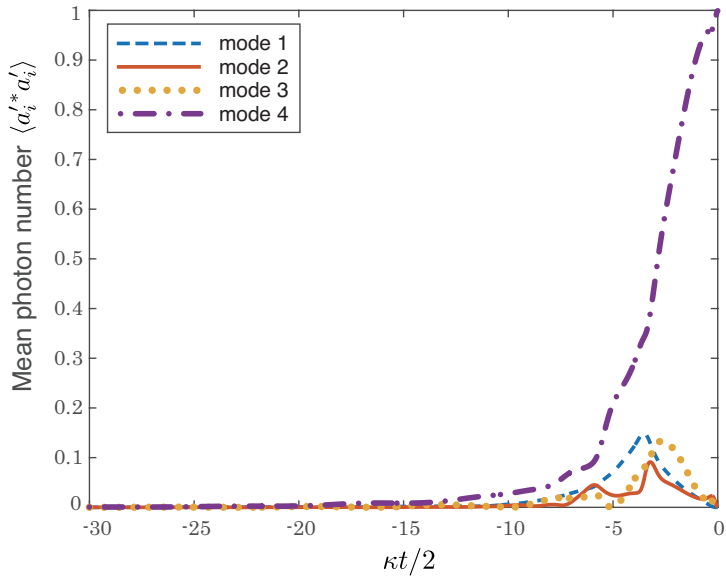}
\caption{
The mean photon number in each mode. 
At time $t_1=0$, the single photon is completely transferred into the mode 
$a'_4$ (purple dot-dashed line).
}
\label{fig:statdynamics}
\end{figure}

Finally, the time-evolution of the mean photon number for each mode is shown in 
Fig.~\ref{fig:statdynamics}, illustrating that certainly the single-photon of the 
input field is transferred perfectly into the target memory mode $a'_4$. 
Note that we can transfer an arbitrarily photon superposition state 
$p_1\ket{1_{\gamma_1}}+p_2\ket{1_{\gamma_2}}$ where $\gamma_1(t)$ 
and $\gamma_2(t)$ are orthogonal functions and $p_i\in{\mathbb C}$ the 
coefficient; 
in this case the transferred state is 
$p_1\ket{1}_3\ket{0}_4+p_2\ket{0}_3\ket{1}_4$ where $\ket{\bullet}_k$ 
is the state corresponding to $a'_k$; see \cite{NYJ}.

\section{Conclusion}

In this paper, we have formulated the optimal control problem that computes 
the best pair of a low-complexity unimodal input pulse function of a single-photon 
state and the corresponding control signal, which achieves the perfect state transfer. 
The numerical simulations demonstrated that, thanks to the aid of time-varying 
control signal, a unimodal input pulse function is obtained, which is simpler 
than the generalized rising exponential function obtained when a constant 
control is employed. 
We again note that the advantage of this method, which simplifies the pulse 
function at the expense of shaping the control signal, relies on the fact that 
implementing a time-varying control signal is in general more feasible than 
the task for shaping a complex waveform of a single-photon, and the fact that 
a flexible tuning of a unimodal pulse function is experimentally realizable.

The future work includes verifying the actual shape of a single-photon input 
pulse which is feasible to generate in an experiment. 
We also point out that the analysis of this paper is limited to the case where 
the pulse function and the control signal have the exact form calculated from 
the numerical optimization. 
The performance may change significantly, if some unwanted noise are present, 
and thus the robustness analysis is crucial for practical application.


\ack
This work was supported by JSPS Grant-in-Aid No. 15K06151 
and JST PRESTO.


\section*{Appendix A: Proof of Eq.~\eref{eqn:IS}}
\label{sec:appendix A}

First, multiplying $\ket{\bm{0},1_{\xi}}$ from the right hand side of 
Eq.~\eref{eqn:HL}, we have 
\begin{eqnarray}
\label{system eq appendix}
     \dot{\bm{a}}(t)\ket{\bm{0},1_{\xi}}
           & = A(t)\bm{a}(t)\ket{\bm{0},1_{\xi}} -C^{\dagger}b(t)\ket{\bm{0},1_{\xi}},
\nonumber\\
     \tilde{b}(t)\ket{\bm{0},1_{\xi}}
           & = C\bm{a}(t)\ket{\bm{0},1_{\xi}}
                   +b(t)\ket{\bm{0},1_{\xi}},
\end{eqnarray}
where $\bm{a}(t)\ket{\bm{0},1_{\xi}}$ denotes the vector of state vectors 
$a_i(t)\ket{\bm{0},1_{\xi}}$. 
The above differential equation can be solved explicitly as follows. 
First let us define the {\it transition matrix} 
$\Phi(t,t_0)=\overleftarrow{{\rm exp}}[\int_{t_0}^t A(s)ds]$, where 
$\overleftarrow{{\rm exp}}$ denotes the time-ordered exponential. 
This satisfies the following properties: 
\[
      \dot{\Phi}(t,t_0)=A(t)\Phi(t,t_0),~~~
      \Phi(t_2, t_1)\Phi(t_1,t_0)=\Phi(t_2,t_0),~~~
      \Phi(t, t)=I. 
\]
Note that $\Phi(t_2, t_1)\Phi(t_1,t_2)=I$. 
Now we define 
$\bm{a}'(t)\ket{\bm{0},1_{\xi}}=\Phi(t_0,t)\bm{a}(t)\ket{\bm{0},1_{\xi}}$; 
then from the differential equation in Eq.~\eref{system eq appendix} together 
with the above properties we have that 
$\dot{\bm{a}}'(t)\ket{\bm{0},1_{\xi}}=-\Phi(t_0,t)C^\dagger \xi(t)\ket{\bm{0},0}$, 
where we have applied $b(t)\ket{\bm{0},1_{\xi}}=\xi(t)\ket{\bm{0},0}$. 
The solution of this equation is, in terms of the original variable 
$\bm{a}(t)\ket{\bm{0},1_{\xi}}$, given by 
\begin{eqnarray}
      \Phi(t_0,t)\bm{a}(t)\ket{\bm{0},1_{\xi}} 
       &= \bm{a}(t_0)\ket{\bm{0},1_{\xi}}
           -\int_{t_0}^t \Phi(t_0,s)C^\dagger \xi(s)ds \ket{\bm{0},0}
\nonumber\\
       &=  -\int_{t_0}^t \Phi(t_0,s)C^\dagger \xi(s)ds \ket{\bm{0},0}, 
\nonumber
\end{eqnarray}
which further yields 
$\bm{a}(t)\ket{\bm{0},1_{\xi}}
=-\int_{t_0}^t \Phi(t,s)C^\dagger \xi(s)ds \ket{\bm{0},0}$. 
Hence the output equation 
$\tilde{b}(t)\ket{\bm{0},1_{\xi}} 
= C\bm{a}(t)\ket{\bm{0},1_{\xi}} + \xi(t)\ket{\bm{0},0}$ in 
Eq.~\eref{system eq appendix} can be expressed as 
\[
     \tilde{b}(t)\ket{\bm{0},1_{\xi}}
      = \Big[ \xi(t) -C\int_{t_0}^t \Phi(t,s)C^\dagger \xi(s)ds \Big] \ket{\bm{0},0}. 
\]
As a consequence the mean photon number of the output field is given by 
\[
    |\tilde{\xi}(t)|^2 
        = \bra{\bm{0},1_{\xi}}\tilde{b}^*(t)\tilde{b}(t)\ket{\bm{0},1_{\xi}}
        = \Big| \xi(t) -C\int_{t_0}^t \Phi(t,s)C^\dagger \xi(s)ds \Big|^2. 
\]
Because the phase change in the pulse function can be made irrelevant, 
we end up with 
\[
     \tilde{\xi}(t)
        = \xi(t) -C\int_{t_0}^t \Phi(t,s)C^\dagger \xi(s)ds. 
\]
Let us now define 
\begin{equation}
\label{def of eta}
     \bm{\eta}(t) 
        = \bra{\bm{0},0}\bm{a}(t)\ket{\bm{0},1_{\xi}}
        = -\int_{t_0}^t \Phi(t,s)C^\dagger \xi(s)ds.
\end{equation}
It is then immediate to see that $\bm{\eta}(t)$ satisfies Eq.~\eref{eqn:IS}.

Before closing this section, let us consider the meaning of $\bm{\eta}(t)$. 
From the definition 
$\bm{a}(t)=[U^*(t) a_1(t_0) U(t), \ldots, U^*(t) a_n(t_0) U(t)]^T$, where $U(t)$ 
is the joint unitary evolution on the system and the field, we have 
$\eta_i(t)=\bra{\bm{0},0}a_i(t)\ket{\bm{0},1_{\xi}}
=\bra{\bm{0},0}U^*(t) a_i(t_0) U(t)\ket{\bm{0},1_{\xi}}$. 
Now define $\ket{\Psi(t)}=U(t)\ket{\bm{0},1_{\xi}}$, which is the joint 
system-field state in the Schr\"{o}dinger picture. 
Also note that, for the passive system, $U(t)\ket{\bm{0},0}=\ket{\bm{0},0}$ holds. 
Therefore, $\eta_i(t)$ can be expressed as 
$\eta_i(t)=\bra{\bm{0},0}a_i(t_0)\ket{\Psi(t)}=\pro{1^{(i)},0}{\Psi(t)}$, 
where $\ket{1^{(i)}}=\ket{0,\ldots, 1, \ldots, 0}$ with $1$ appearing only in 
the $i$th entry. 
This means that $\eta_i(t)$ represents how much the $i$th system mode 
is excited at time $t$, by the single-photon driving. 
In particular, if the single photon field state is completely transferred to the 
system at time $t_1$ and the whole system-field state gains the form 
$\ket{\Psi(t_1)}=\sum_i \lambda_i \ket{1^{(i)}}\ket{0}$, then we have 
$\eta_i(t_1)=\lambda_i$. 
Thus, $\eta(t_1)$ coincides with the superposition coefficient of the system state 
when the perfect state transfer is completed.


\section*{Appendix B: Optimization algorithm}
\label{sec:appendix B}

Here we present the procedure for solving the optimization problem studied 
in this paper.


\subsection{Euler-Lagrange equation}

Consider the following real-valued deterministic system: 
\begin{eqnarray}
        \dot{x}(t)=f(x(t),u(t),t),~~~x(t_i)=x_{t_i},
\label{eqn:function}
\end{eqnarray}
where $x(t)$ is a vector of variables, $f(x,u,t)$ is a vector of functions, and 
$u(t)$ is the control signal. 
$x(t_i)=x_{t_i}$ is a fixed initial state at the initial time $t_i$. 
The problem is to find the optimal $u(t)$ that minimizes the cost function
\begin{eqnarray}
          J[u]=\varphi(x(t_f))+\int_{t_i}^{t_f}L(x(t),u(t),t)dt,
\end{eqnarray}
where $\varphi$ and $L$ are scalar-valued functions defined in $[t_i, t_f]$. 
To solve the problem, we utilize the variational method and aim to find the 
minimum of the following functional:
\begin{eqnarray}
        \bar{J}=\varphi(x(t_f))+\int_{t_i}^{t_f}\left(H(x,u,p,t)-p^T\dot{x}\right)dt,
\end{eqnarray}
where $p(t)$ is a vector of Lagrange multipliers associated with 
Eq.~\eref{eqn:function}, and
\begin{eqnarray}
        H(x,u,p,t)=L(x,u,t)+p^Tf(x,u,t)
\label{eqn:HamiltonF}
\end{eqnarray}
is the Hamilton function. 
The variation $\delta\bar{J}$ is calculated as 
\begin{eqnarray}
	\delta\bar{J}
	    &=\frac{\partial\varphi}{\partial x}\delta x(t_f)
	      +\int_{t_i}^{t_f}\left(\frac{\partial H}{\partial x}\delta x
	      +\frac{\partial H}{\partial u}\delta u-p^T\delta\dot{x}\right)dt
\nonumber\\
            &=\frac{\partial\varphi}{\partial x}\delta x(t_f)
            -\left[p^T\delta x\right]^{t_f}_{t_i}
               +\int_{t_i}^{t_f}\left(\frac{\partial H}{\partial x}\delta x
               +\frac{\partial H}{\partial u}\delta u
               +\dot{p}^T\delta x\right)dt
\nonumber\\
            &=\left(\frac{\partial\varphi}{\partial x}
                       -p^T(t_f)\right)\delta x(t_f)
               +\int_{t_i}^{t_f}\left\{\left(\frac{\partial H}{\partial x}
               +\dot{p}\right)\delta x
               +\frac{\partial H}{\partial u}\delta u\right\}dt,
\nonumber
\end{eqnarray}
where we have used $\delta x(t_i)=0$ since $x(t_i)$ is fixed. 
The stationary condition of the functional is thus
\begin{eqnarray}
\label{eqn:fx}
           &\dot{x}(t)=f(x(t),u(t),t),~~~x(t_i)=x_{t_i},
\\
\label{eqn:adjfx}
           &\dot{p}(t)=-\left(\frac{\partial H}{\partial x}\right)^T(x,u,p,t),~~~
           p(t_f)=\left(\frac{\partial \varphi}{\partial x}\right)^T(x(t_f)),
\\
\label{eqn:dHdu}
           &\frac{\partial H}{\partial u}(x,u,p,t)=0, 
\end{eqnarray}
which are called the Euler-Lagrange equations.

Now, to apply the above Euler-Lagrange method to our case, we represent 
the complex-valued dynamics \eref{eqn:subjectsimple} as the real-valued 
one as follows: 
\begin{eqnarray}
\label{eqn:subjectto}
        \frac{d}{dt}\left[\begin{array}{c}
                    \bm{x}_R\\
                    \bm{x}_I
                \end{array}\right]
          = \left[\begin{array}{cc}
                    -A_{0R}-A_{1R}u & A_{0I}+A_{1I}u\\
                    -A_{0I}-A_{1I}u & -A_{0R}-A_{1R}u
                \end{array}\right]
                \left[\begin{array}{c}
                    \bm{x}_R \\
                    \bm{x}_I
                \end{array}\right],
\end{eqnarray}
where the subscripts $\bullet_R$ and $\bullet_I$ denote the real and the 
imaginary component of the matrix or the vector (recall 
$\bm{x}=[\xi, \bm{\eta}^T_1, \bm{\eta}^T_2]^T$). 
Also, the Hamilton function is now given by 
\begin{eqnarray}
        H(\bm{x},u,p,t)
               &=\alpha \, {\rm exp}\left(2h(t)
                        \left\{\xi_R\dot{\xi}_R+\xi_I\dot{\xi}_I\right\}\right)
                           +\beta u^2
\nonumber\\
         & \hspace{-3em} +\bm{p}^T
                   \left[\begin{array}{cc}
                      -A_{0R}-A_{1R}u & A_{0I}+A_{1I}u \\
                      -A_{0I}-A_{1I}u & -A_{0R}-A_{1R}u
                   \end{array}\right]
                 \left[\begin{array}{c}
                    \bm{x}_R \\
                    \bm{x}_I
                \end{array}\right].
\nonumber
\end{eqnarray}
%


\subsection{Steepest gradient method}

We utilize the following procedure \cite{Bryson Ho} for determining the optimal 
control $u(t)$:
\begin{enumerate}
\item Prepare the initial control $u(t)$ defined in $[t_i, t_f]$.

\item Calculate the backward solution $x(t)$ subjected to Eq.~\eref{eqn:fx}, 
starting from $t_f$. \label{enum:x}

\item Using the solution obtained in the step \eref{enum:x}, calculate the forward 
solution $p(t)$ subjected to Eq.~\eref{eqn:adjfx}, starting from $t_i$. 

\item If $\left(\int_{t_i}^{t_f}\|\partial H/\partial u\|^2dt\right)^{1/2}$ 
is small enough, terminate. Otherwise, go to the next step. 

\item Let $s=-\left(\partial H/\partial u\right)^T$ and search $\epsilon>0$ 
such that the cost function $J[u+\epsilon s]$ is minimal. 
Then replace $u+\epsilon s$ by $u$, and go to the step \eref{enum:x}.
\label{enum:alpha}
\end{enumerate}
The procedure for finding the optimal $\epsilon$ in the step \eref{enum:alpha} 
can be conducted by the line search algorithm. 
If the precise solution is not required, the following Wolfe conditions \cite{Nocedal} 
are convenient to use: 
\begin{description}
         \item[Armijo rule:] $J[u+\epsilon s]\leq J[u]+\mu_1 J'[u]\epsilon$, 
         \item[Curvature condition:] $J'[u+\epsilon s]\geq\mu_2 J'[u]$, 
\end{description}
where $J'[u+\epsilon s]=\int_{t_i}^{t_f}\frac{\partial L(x,u+\epsilon s,t)}
{\partial \epsilon}dt$ is the derivative of the cost functional $J$ with respect to 
$\epsilon$. 
If $\epsilon_{\rm opt}$ satisfies the above conditions for some constants 
$0<\mu_1<\mu_2<1$, one may assume that the function 
$J[u+\epsilon_{\rm opt} s]$ has decreased sufficiently and take 
$\epsilon_{\rm opt}$ as the solution in the step \eref{enum:alpha}.



\section*{References}


\begin{thebibliography}{10}

\bibitem{DiVincenzo}
D. P. Divincenzo, 
The physical implementation of quantum computation, 
Fortschr. Phys. \textbf{48}, 9-11, 771/783 (2000).

\bibitem{Phillips}
D. F. Phillips, A. Fleischhauer, A. Mair, R. L. Walsworth, and M. D. Lukin, 
Storage of light in atomic vapor, 
Phys. Rev. Lett. \textbf{86}, 783 (2001).

\bibitem{Julsgaard}
B. Julsgaard, J. Sherson, J. I. Cirac, J Fiurasek, and E. S. Polzik, 
Experimental demonstration of quantum memory for light, 
Nature \textbf{432}, 482, (2004).

\bibitem{Chang}
D. E. Chang, A. H. Safavi-Naeini, M. Hafezi, and O. Painter, 
Slowing and stopping light using an optomechanical crystal array, 
New J. Phys. \textbf{13}, 023003 (2011).

\bibitem{Bao}
X. Bao, A. Reingruber, P. Dietrich, J. Rui, A. D\"{u}ck, T. Strassel, L. Li, 
N. Liu, B. Zhao, and J. Pan, 
Efficient and long-lived quantum memory with cold atoms inside a ring cavity, 
Nature Physics \textbf{8}, 517/521 (2012).





\bibitem{Fleischhauer}
M. Fleischhauer and M. D. Lukin, 
Dark-state polaritons in electromagnetically induced transparency, 
Phys. Rev. Lett. \textbf{84}, 5094 (2000).

\bibitem{Lvovsky}
A. I. Lvovsky, B. C. Sanders, and W. Tittel, 
Optical quantum memory, 
Nature Photonics \textbf{3}, 706 (2009). 

\bibitem{Bussieres}
F. Bussi\`{e}res, N. Sangouard, M. Afzelius, H. de Riedmatten, C. Simon, 
and W. Tittel, 
Prospective applications of optical quantum memories, 
Journal of Modern Optics \textbf{60}, 1519/1537 (2013).


\bibitem{Gorshkov2007}
A. V. Gorshkov, A. Andre, M, Fleischhauer, A. S. S\o rensen, and M. D. Lukin, 
Universal approach to optimal photon storage in atomic media, 
Phys. Rev. Lett. \textbf{98}, 123601 (2007).

\bibitem{Novikova}
I. Novikova, A. V. Gorshkov, D. F. Phillips, A. S. S\o rensen, M. D. Lukin, 
and R. L. Walsworth, 
Optimal control of light pulse storage and retrieval, 
Phys. Rev. Lett. \textbf{98}, 243602 (2007).

\bibitem{Gorshkov}
A. V. Gorshkov, T. Calarco, M. D. Lukin, and A. S. S\o rensen, 
Photon storage in $\Lambda$-type optically dense atomic media. 
IV. Optimal control using gradient ascent, 
Phys. Rev. A \textbf{77}, 043806 (2008).

\bibitem{Phillips 2008}
N. B. Phillips, A. V. Gorshkov, and I. Novikova, 
Optimal light storage in atomic vapor, 
Phys. Rev. A \textbf{78}, 023801 (2008). 

\bibitem{Novikova 2008}
I. Novikova, N. B. Phillips, and A. V. Gorshkov, 
Optimal light storage with full pulse-shape control, 
Phys. Rev. A \textbf{78}, 021802 (2008).



\bibitem{Gough 2008}
J. Gough, R. Gohm, and M. Yanagisawa, 
Linear quantum feedback networks, 
Phys. Rev. A \textbf{78}, 062104 (2008). 

\bibitem{Guta 2016}
M. Guta and N. Yamamoto, 
System identification for passive linear quantum systems, 
IEEE Trans. Automat. Contr. \textbf{61}-4, 921/936 (2016). 


\bibitem{NYJ}
N. Yamamoto and M. R. James, 
Zero-dynamics principle for perfect quantum memory in linear networks, 
New J. Phys. \textbf{16}, 073032 (2014).




\bibitem{Du 2012}
S. Zhang, C. Liu, S. Zhou, C.-S. Chuu, M. M. T. Loy, and S. Du, 
Coherent control of single-photon absorption and reemission in a two-level 
atomic ensemble, 
Phys. Rev. Lett. \textbf{109}, 263601 (2012).

\bibitem{Gulati PRL}
B. Srivathsan, G. K. Gulati, A. Cere, B. Chng, and C. Kurtsiefer, 
Reversing the temporal envelope of a heralded single photon using a cavity, 
Phys. Rev. Lett. \textbf{113}, 163601 (2014).

\bibitem{Gulati PRA}
G. K. Gulati, B. Srivathsan, B. Chng, A. Cere, D. Matsukevich, and C. Kurtsiefer, 
Generation of an exponentially rising single-photon field from parametric 
conversion in atoms, 
Phys. Rev. A \textbf{90}, 033819 (2014). 

\bibitem{Lvovsky 2015}
Z. Qin, A. S. Prasad, T. Brannan, A. MacRae, A. Lezama, and A. I. Lvovsky, 
Complete temporal characterization of a single photon, 
Light Sci. Appl. \textbf{4}, e298 (2015).
 
\bibitem{Ogawa}
H. Ogawa, H. Ohdan, K. Miyata, M. Taguchi, K. Makino, H. Yonezawa, 
J. Yoshikawa, and A. Furusawa, 
Real-time quadrature measurement of a single-photon wave packet 
with continuous temporal-mode matching, 
Phys. Rev. Lett. \textbf{116}, 233602 (2016).



\bibitem{Werschnik}
J. Werschnik and E. K. U. Gross, 
Quantum optimal control theory, 
J. Phys. B: At. Mol. Opt. Phys. \textbf{40}, 175 (2007). 

\bibitem{Cong}
S. Cong, 
{\it Control of Quantum Systems: Theory and Methods} 
(Wiley, Singapore, 2014). 



\bibitem{Gardiner}
C. W. Gardiner and P. Zoller, 
\textit{Quantm Noise} 
(Springer-Verlag, Berlin, 2004).

\bibitem{Hush}
M. R. Hush, A. R. R. Carvalho, M. Hedges, and M. R. James, 
Analysis of the operation of gradient echo memories using a quantum 
input-output model, 
New J. Phys. \textbf{15}, 085020 (2013).

\bibitem{Zhang}
G. Zhang and M. R. James, 
On the response of quantum linear systems to single photon input fields, 
IEEE Trans. Autom. Control \textbf{58}, 1221 (2013). 




\bibitem{Leuchs 2000}
S. Quabis, R. Dorn, M. Eberler, O. Glockl, and G. Leuchs, 
Focusing light to a tighter spot, 
Optics Communications \textbf{179}, 1 (2000). 

\bibitem{Bader}
M. Bader, S. Heugel, A. L. Chekhov, M. Sondermann, and G. Leuchs, 
Efficient coupling to an optical resonator by exploiting time-reversal symmetry, 
New J. Phys. \textbf{15} 123008 (2013). 




\bibitem{Rempe2002}
A. Kuhn, M. Hennrich, and G. Rempe, 
Deterministic single-photon source for distributed quantum networking, 
Phys. Rev. Lett. \textbf{89}, 067901 (2002).

\bibitem{Kuhn2011}
P. B. R. Nisbet-Jones, J. Dilley, D. Ljunggren, and A. Kuhn, 
Highly efficient source for indistinguishable single photons of controlled shape, 
New J. Phys. \textbf{13} 103036 (2011). 

\bibitem{Rempe2012}
S. Ritter, C. Nolleke, C. Hahn, A. Reiserer, A. Neuzner, M. Uphoff, 
M. Mucke, E. Figueroa, J. Bochmann, and G. Rempe, 
An elementary quantum network of single atoms in optical cavities, 
Nature \textbf{484}, 195 (2012).

\bibitem{Hayasaka2004}
M. Keller, B. Lange, K. Hayasaka, W. Lange, and H. Walther, 
Continuous generation of single photons with controlled waveform 
in an ion-trap cavity system, 
Nature \textbf{431}, 1075 (2004). 

\bibitem{Filipp2014}
M. Pechal, L. Huthmacher, C. Eichler, S. Zeytinolu, A. A. Abdumalikov, 
S. Berger, A. Wallraff, and S. Filipp, 
Microwave-controlled generation of shaped single photons in circuit quantum 
electrodynamics, 
Phys. Rev. X \textbf{4}, 041010 (2014). 

\bibitem{Harris2008}
P. Kolchin, C. Belthangady, S. Du, G. Y. Yin, and S. E. Harris, 
Electro-optic modulation of single photons, 
Phys. Rev. Lett. \textbf{101}, 103601 (2008).

\bibitem{Riedmat2016}
P. Farrera, G. Heinze, B. Albrecht, M. Ho, M. Chavez, C. Teo, 
N. Sangouard, and H. de Riedmatten, 
Generation of single photons with highly tunable wave shape from a cold 
atomic quantum memory, 
Nature Commun. \textbf{7} 13556 (2016). 




\bibitem{Duan}
L. M. Duan, J. I. Cirac, and P. Zoller, 
Three-dimensional theory for interaction between atomic ensembles
and free-space light, 
Phys. Rev. A \textbf{66}, 023818 (2002). 

\bibitem{Parkins}
A. S. Parkins, E. Solano, and J. I. Cirac, 
Unconditional two-mode squeezing of separated atomic ensembles, 
Phys. Rev. Lett. \textbf{96}, 053602 (2006). 

\bibitem{Ficek}
G. Li, S. Ke, and Z. Ficek, 
Generation of pure continuous-variable entangled cluster states of four separate 
atomic ensembles in a ring cavity, 
Phys. Rev. A \textbf{79}, 033827 (2009). 

\bibitem{Bryson Ho}
A. E. Bryson Jr. and Y. C. Ho, 
\textit{Applied Optimal Control} 
(Taylor and Francis, New York, 1975).

\bibitem{Nocedal}
J. Nocedal and S. Wright, 
\textit{Numerical Optimization} 
(Springer, New York, 2006).

\end{thebibliography}
\end{document}